# Modification of electron states in CdTe absorber due to a buffer layer in CdTe/CdS solar cells

Y. G. Fedorenko[1, a)], J. D. Major[1], A. Pressman[1], L. J. Phillips[1], K. Durose[1]

[1]Stephenson Institute for Renewable Energy and Department of Physics, School of Physical Sciences, Chadwick Building, University of Liverpool, Liverpool L69 7ZF, UK

[a)] Electronic mail: y.fedorenko@liverpool.ac.uk

Abstract

By application of the ac admittance spectroscopy method, the defect state energy distributions were determined in CdTe incorporated in thin film solar cell structures concluded on ZnO, ZnSe, and ZnS buffer layers. Together with the Mott-Schottky analysis, the results revealed a strong modification of the defect density of states and the concentration of the uncompensated acceptors as influenced by the choice of the buffer layer. In the solar cells formed on ZnSe and ZnS, the Fermi level and the energy position of the dominant deep trap levels were observed to shift closer to the midgap of CdTe suggesting the mid-gap states may act as recombination centers and impact the open-circuit voltage and the fill factor of the solar cells. For the deeper states, the broadening parameter was observed to increase indicating fluctuations of the charge on a microscopic scale. Such changes can be attributed to the grain-boundary strain and the modification of the charge trapped at the grain-boundary interface states in polycrystalline CdTe.

I. INTRODUCTION

CdTe is a promising and widely investigated semiconductor in thin film photovoltaics.[1] The formation of the CdTe/CdS heterojunction (HJ) and doping of CdTe are considered as major factors which determine the cell efficiency. It is expected that diffusion or mixing would take a place across the CdTe/CdS HJs under the cell processing. The incorporation of impurities predominantly occurs within the grain boundaries (GB) and alters the surface band bending,[2] thereby strongly affecting the electronic properties of polycrystalline solar cells. The band bending towards either inversion or accumulation has been considered beneficial for the CdTe solar cell operation since electrons (at the inversion band bending) or holes (at the accumulation band bending) would be repelled from the GB reducing the recombination rate and enhancing the charge carrier collection along the grain boundaries.[3,4,5] The incorporation of Cl in polycrystalline CdTe has been suggested to induce hole conductivity in CdTe due to the formation of $\beta$ acceptor complexes which decay into chlorine donors and A-centers when chlorine content decreases.[6] In single crystal CdTe:Cl, the defect clustering can play a role in the compensation processes giving rise to the formation of a wide energy spectrum of deep trap levels within $E_V + (0.15 - 0.9)$ eV.[7] The main compensation processes are similar in both polycrystalline and single crystal CdTe:Cl, although additional contribution may stem from Cl-related defects in the vicinity of extended defects.[8] Positron annihilation experiments performed on polycrystalline CdTe films confirmed the stabilizing effect of Cl on the divacancy $V_{Cd}$-$V_{Te}$ defect formation.[9]

In superstrate solar cell design a buffer layer deposited on a conductive transparent electrode is often used to enhance the open circuit voltage and the fill factor, prevent the appearance of a potential barrier at the interface with conductive oxides[10] or induce charge redistribution in a solar cell via creation of defects in a CdS layer.[11] In the latter case, a solar cell comprising a conductive transparent electrode and a CdTe/CdS HJ can function as a metal-oxide-semiconductor (MOS) structure due to a significant density of acceptor traps in CdS. Therefore, using different buffer layers one can expect to alter electronic properties of CdS and change the electrostatic conditions at the CdTe/CdS heterojunction. This prompted us to conduct investigation of the energy distribution of defect-related electron states in CdTe as influenced by the choice of a buffer layer.

In this work we show that a buffer layer in CdTe/CdS heterojunction solar cells can impact the electronic properties of the CdTe/CdS junction and the CdTe absorber layer. Electron density of states (DOS) are compared in CdTe solar cells grown on ZnO, ZnSe, and ZnS buffer layers. In respect to the CdTe/CdS/ZnO solar cells, the structures concluded on ZnS and ZnSe revealed a significantly reduced doping level in the CdTe and an increased density of the electrically active defects whose energy levels were shifted towards the midgap.

## II. EXPERIMENT

The samples for this study were prepared according to a procedure similar to that described elsewhere.[12] The structures of CdTe/CdS solar cells were identical except for introduction of different buffer layers, ZnO, 120nm thick, ZnS, 110 nm thick, and ZnSe, 100 nm thick grown by magnetron sputtering on TEC 7 glass cleaned in solvents and deionised water. The CdS films of 200 nm were deposited above by magnetron sputtering at 200°C. CdTe films were grown using close space sublimation (CSS) at $T_{source}$ = 605°C and $T_{sub}$ = 520°C. The layer thickness of CdTe films was in the range of 3.5-4.0 μm. The nitric-phosphoric (NP) etch was applied prior to MgCl$_2$ treatment which was done by placing the samples in a tube furnace at 430°C in air ambient for 25-30 minutes. Prior to the back contact deposition the cells were etched in the NP solution for 30 s. The averaged parameters of each series are compiled in Table 1.

The solar cells were characterised using capacitance-voltage (*C–V*) and ac admittance measurements (AS) carried out in the dark (Solartron 1260 Impedance Analyzer equipped with a 1296 Dielectric Interface). In order to extract the trap energy distribution $N_t(E)$ from the capacitance-frequency response the analytical model for *p- i- n* junction proposed by Walter *et al.* [13] was used. In this analysis, the trap density at the energy scale is related to the derivative of the capacitance with respect to the frequency as $N_t(E) = -\frac{V_{bi}}{qW_D}\frac{dC}{d\omega}\frac{\omega}{kT}$, where $V_{bi}$ is the built-in voltage and $W_D$ is the width of the space charge region. The energy scale is linked to the frequency as $E_\omega = kT \ln \frac{2\beta_{p,n} N_{v,c}}{\omega}$, were $\beta_{p,n}$ are the capture coefficients for holes and electrons, and $N_{v,c}$ is the effective density of states in the valence or conduction band.

Charge carrier freeze-out effects were not detected in the temperature range corresponding to the deep trap ac responses as evidenced by the *C-V* data shown in Supplemental Information (SI), Fig. 1(a, b).[14] The activation energy of the series resistance for the studied samples deduced using the differential method[15] was within 160-170 meV in the temperature range, 250−350 K, for the samples grown on ZnS and ZnSe, SI, Fig. 2.[14] The back contact barrier height derived from the temperature dependence of the saturation current $J_t − (1/T)$ was about 0.5 eV, SI, Fig. 3.

III.  RESULTS AND DISCUSSION

The Mott-Schottky plots for the samples grown on ZnO, ZnSe, and ZnS buffer layers and distribution of the apparent carrier density in the absorber are shown in Fig. 1 (a) and Fig. (b), respectively. The depth value given in Fig. 1 (b) is normalized to the depletion width $W_D$ calculated from the p-n junction capacitance $W_D = \varepsilon\varepsilon_0 A/C$, and measured from the CdTe/CdS interface. In order to exclude the back contact influence, the data were considered at the bias V < 0.5 V. The cells comprising ZnS or ZnSe as a buffer layer revealed lower values of the built-in voltage $V_{bi}$ being decreased to 0.54 V and 0.32 V, respectively, indicating a smaller barrier height at the heterojunction. The effect of the buffer layer on the background doping concentration $\rho$ is shown in Fig. 1(b). The obtained values of $\rho$ throughout the CdTe thickness are very close in both samples, ZnS and ZnSe, and they are lower by approx. half of an order of magnitude than that of the cells grown on ZnO. The data presented in Fig. 1 (b) indicate that ZnS or ZnSe buffer layers inhibit the electrical activation by Cl atoms in CdTe. The shape of the doping profiles for the cells deposited on the ZnO buffer layer is similar to those reported earlier.[16] It is worth noting that similar non-uniform distributions of the net ionized impurity concentration were observed in depletion region of the junction when the chlorine treatment was implemented in NaCl, KCl and MnCl$_2$.[12] In the latter cases, the lower doping efficiency has been linked to the cation charge state which differs from $2^+$ in chlorides other than MgCl$_2$ or CdCl$_2$. The higher carrier density towards the CdTe/CdS interface observed in Fig. 1 (b) has been often ascribed to the presence of deep levels in the junction.[16] The apparent carrier density in the vicinity of the back contact seems to saturate at the level of $(0.7 - 2.0) \times 10^{15}$ cm$^{-3}$ regardless of the chosen buffer layer indicating that no significant difference in the back contact barrier height of the studied solar cells could be expected.

The energy distributions of traps in the CdTe/CdS junction are presented in Fig. 2. Only one deep defect level is observed in the studied samples in accordance to earlier findings on CdTe/CdS solar cells subjected to the NP etch prior to evaporation of Au contacts.[17] A significant energy shift of the activation energy from 0.38 eV (Fig. 2(a)), to 0.5 eV (Fig. 2 (b)), and to 0.75-0.76 eV (Fig. 2 (c)), was observed for the CdTe/CdS cells grown on ZnO, ZnSe, and ZnS buffer layers, respectively. The Arrhenius plots are shown in SI, Fig. 4.

One can argue that the back contact can modify the DOS spectra. This is usually observed if the near-contact region can retain a sufficient quantity of charge to form a dipole which modulates the barrier height. The conditions to observe admittances from the back Au electrode imply that the fast charge transfer via tunnelling is blocked. Such an influence is expected at frequencies higher than 50 kHz in CdTe/CdS samples,[18] and it is not consistent with the low-frequency responses observed here for the CdTe/CdS/ZnS and CdTe/CdS/ ZnSe structures. Also, an example of the capacitance derivative spectra taken under the reverse bias can evidence that the AS responses stem from the CdTe/CdS junction, not from the Au/CdTe contact (SI, Fig.5).[14] Finally, the trap energy positions and emission rates for the deep trap levels correspond to the well-studied electrically active defects identified in the energy range from 0.35 eV to 0.75 eV in differently processed CdTe/CdS solar cells.[19] The broadening parameter in the inset of Fig. 2 (a) indicates that the effects of disorder are significant in polycrystalline CdTe:Cl films. The broadening of DOS can be modelled in terms of potential fluctuations caused by the random distribution of doping impurities or defects. Randomly distributed dopants or defect clusters can lead to unavoidable fluctuations of the charge on a microscopic scale resulting in potential fluctuations. The disorder parameter was found to vary from 26 meV to 100 meV in the temperature range 180−400 K [the inset in Fig.2(a)]. Similar values have been typically found in polymer semiconductors, although in that case the broadening of the DOS is due to the energy disorder mediated by the Coulomb interactions.[20]

Bias dependence of AS responses is an inherent property of the interface states. When AS responses are not influenced by the bias this is a primary indication that the Fermi level has been pinned by the interface traps. For instance, the Fermi level pinning at the CdS/CuInSe$_2$ interface by shallow acceptor levels has been observed in Ref. 21. However, the Fermi level pinning at the bulk defect density of $10^{15}$ cm$^{-3}$−$10^{16}$ cm$^{-3}$ observed in the CdTe/CdS solar cells studied in our work is unlikely. We suppose that the electrically active defects in CdTe appear due to bulk traps rather than to the interface states in the CdTe/CdS junction. In conjunction with the lower net carrier concentration found in CdTe when ZnS and ZnSe buffer layers are employed, the energy shift in the DOS distributions and the lower doping density could be attributed to the lack of donor-acceptor complexes, which promote p-type doping in CdTe. Instead, point defect clusters which contribute to Shockley-Read-Hall recombination could be formed causing a decrease of the open circuit voltage and the fill factor of solar cells.[22] It can be speculated that when the chlorine distribution is *locally* non-uniform, for instance, as a result of strain this could either modify distribution of $V_{Cd}$ defects and cause fluctuations of the $V_{Cd}$ concentration in the vicinity of an interstitial atom which is involved in the formation of the compensating defects, or locally increase concentration of Cl, a donor impurity in CdTe. Since Te-rich surfaces have been reported to experience a tensile stress,[23] this explanation is not unlikely. Also, the Te-rich grain boundaries in CdTe has been found to differ in electrical activity of the deep trap levels as modified by the duration of the NP etch.[17] The longer etching times in the NP solution preferentially removed AS responses stemming from the shallower acceptor levels in CdTe suggesting that stronger bonding in the defects responsible for the deep trap levels might be linked to the Cd vacancy defect clustering. Furthermore, the increased conductance and noisy spectra of the capacitance derivative in Fig. 5 of Ref. 17 could indicate the increased surface roughness at the Te-rich GBs. The examples of modification of the electronic properties of solids experiencing the lattice expansion or contraction are numerous. In crystalline semiconductor heterojunctions and superlattices the effects of strain are utilized to tailor the electronic band structure. Here, we note that the effect of strain called as the *edge defeat* effect impedes passivation of silicon dangling bonds by molecular hydrogen in the $Si/SiO_2$ interface due to a large spread in the activation energy of passivation/depassivation reactions.[24] Pertaining to polycrystalline CdTe, anneals in the presence of chlorine has been found to vary the activation energy of the acceptor trap as a result of a valence-band deformation at grain boundaries.[25] The Cl-treatment of the solar cells in work [25] was implemented using different amount of the Cl-containing reagent to anneal each particular solar cell. The CdS/CdTe solar cells studied in our work were annealed using the same quantity of MgCl$_2$ solution applied to CdTe in excess. Therefore, the concentration-limited supply of chlorine is unlikely. It is plausible to suggest that passivation of the Te core by Cl in polycrystalline CdTe[26] might be hindered in the conditions of strain.

Figure 3 exemplifies *I-V* characteristics taken on the CdTe/CdS cells deposited on ZnO, Fig. 3 (a, b). The charge carrier transport mechanism at both the forward, Fig 3(a), and the reverse, Fig. 3(b), bias is space-charge-limited conduction (SCLC), the characteristic property of compensated semiconductors, as it is evidenced by the slope of the *log (I)-log(V)* curves increasing from 1 to 1.4-1.6 in the studied temperature range. The data can be fitted by the expression $I \propto V^m$ with *m* ranging from 1.4 to 1.6 in the temperature range from 300 K to 110 K corresponding to monoenergetic trap levels. The steep increase in *IV* curves, characteristic to the trap filling regime is not observed suggesting that the traps are too deep to influence the charge carrier transport in the film. The relatively high ohmic current at low voltages is most probably caused by impurities in the utilized semiconductors. Using the expression for the critical voltage $V_T = qN_td^2/2\varepsilon\varepsilon_0$, were *q* is the elemental charge, $N_t$ is the trap density, *d* is the film thickness, $\varepsilon$ is the film permittivity, and $\varepsilon_0$ is the permittivity of the free space. Given that the relative permittivity for CdTe is 10.16, and for CdS is 10 the trap density $N_t$ can be estimated as $\sim 7 \cdot 10^{15}$ cm$^{-3}$ and $\sim 4.2 \cdot 10^{16}$ cm$^{-3}$ for positive and negative polarity of the applied bias, respectively. It is apparent that the rectification ratio in the CdTe/CdS junction is dependent on the trap filling under positive and negative bias.

SCLC was not observed when the cell structures were grown on ZnS or ZnSe layer as shown in Fig. 4. Possible injection and bulk-limited conduction mechanisms in solar cells based on ZnS and ZnSe buffer layers are analysed by plotting the *IV*-curves in the Schottky, Pool-Frenkel, and the Fowler–Nordheim (FN) coordinates, Fig. 4. The Poole-Frenkel conduction for internal-field-assisted thermal ionization or the Schottky over barrier emission are operative at higher temperature and electric field. To discriminate between the bulk and the barrier limited mechanism, the relative densities of donor and acceptor traps in CdTe must be known. Alternatively, conductive electrodes having different work functions could be employed. Nevertheless, the barrier conduction mechanism can be inferred from the FN plots, Fig. 4(c). A linear dependence of the currents in the higher field is typical for electron tunnelling.[27] In the low-field range the currents do not exhibit the FN conduction, suggesting a trap-assisted electron transport. At temperatures below 200 K the FN behaviour is not observed indicating that the mechanism of the tunnel injection is no longer valid. Much lower carrier densities in the diode base than those observed in Fig. 3 for the devices concluded on ZnO buffer layer may explain why rectification in all devices which utilize ZnSe or ZnS buffer layers vanishes in temperature range of 230-200 K. The reverse current is considered as a generation current from the obtained linear relationship $logI$-$V^{1/2}$, Fig. 4(d). Though a buffer layer or a copper-doped CdS in CdTe solar cell has been supposed to function as a dielectric in a MOS-structure due to a strongly increased screening length in ZnO or/and CdS,[11] this explanation which is applicable on a macroscopic scale does not consider complex electronic structure of the GBs that can be modified in the growth of CdTe following the Volmer-Weber mechanism. The nucleation and growth of CdTe may be influenced by the choice of a buffer layer in a solar cell structure.

IV. CONCLUSIONS

By using ac admittance spectroscopy the defect densities of states in CdTe absorber were studied in solar cells deposited on ZnO, ZnS, and ZnSe buffer layers. Compared to the CdTe/CdS solar cells grown on ZnO, the dominant deep trap levels in CdTe were shifted in energy closer to the midgap when ZnS or ZnSe were used as the buffer layers. The deep trap level formation in the cells comprising ZnS or ZnSe is accompanied by a reduction of the doping density in CdTe implying that the Cl activation and the formation of the ionized acceptors are impeded. Such modification of the defect energy position and the density may originate from the valence-band deformation due to strain in the grain boundaries of CdTe. Concomitantly, the charge carrier transport mechanism in the solar cells is changed from the bulk-limited to the barrier-limited.

Figure captions

Table I. Averaged performance parameters of solar cells containing different buffer layers

FIG. 1. (a) $1/C^2$–$V$ plots for samples with different buffer layers. The data were measured at T =300K and frequency of 10 kHz. The inset shows the Mott–Schottky plot for the CdTe/CdS/ZnO solar cell. The straight black lines are the fitting curves to estimate the built-in potential $V_{bi}$; (b) The apparent doping density profiles $\rho(W_D)$ obtained from the $1/C^2$–$V$ plots.

FIG. 2. The density of states in the CdTe/CdS solar cells deposited on different buffer layers: (a) ZnO, (b) ZnSe, (c) ZnS. The broadening parameter derived from the DOS in Figs. (a, b) is shown in the inset of Fig. 2(a). The emission rates were determined from the Arrhenius plots taken at zero bias, SI, Fig. 4.

FIG. 3. Forward (a) and reverse (b) current-voltage characteristics of CdTe/CdS junctions deposited on ZnO. The contact area is 0.24 cm$^2$.

FIG. 4. The forward current in CdTe/CdS/ZnS solar cells in Schottky (a), Poole-Frenkel (b), and the Fowler–Nordheim (c) coordinates; the reverse current in $logI$ – $V^{1/2}$ scale (d).

TABLE I

| Buffer layer | Efficiency (%) | Fill Factor (%) | $J_{SC}$ (mA/cm$^2$) | $V_{OC}$ (V) |
|---|---|---|---|---|
| 120nm ZnO | 8.22 ± 1.70 | 59.25 ± 2.75 | 20.30 ± 1.67 | 0.675 ± 0.060 |
| 110nm ZnS | 0.85 ± 0.29 | 27.61 ± 3.39 | 6.03 ± 1.59 | 0.507 ± 0.024 |
| 100nm ZnSe | 0.98 ± 0.25 | 33.38 ± 1.92 | 6.22 ± 1.33 | 0.468 ± 0.022 |

FIG. 1.

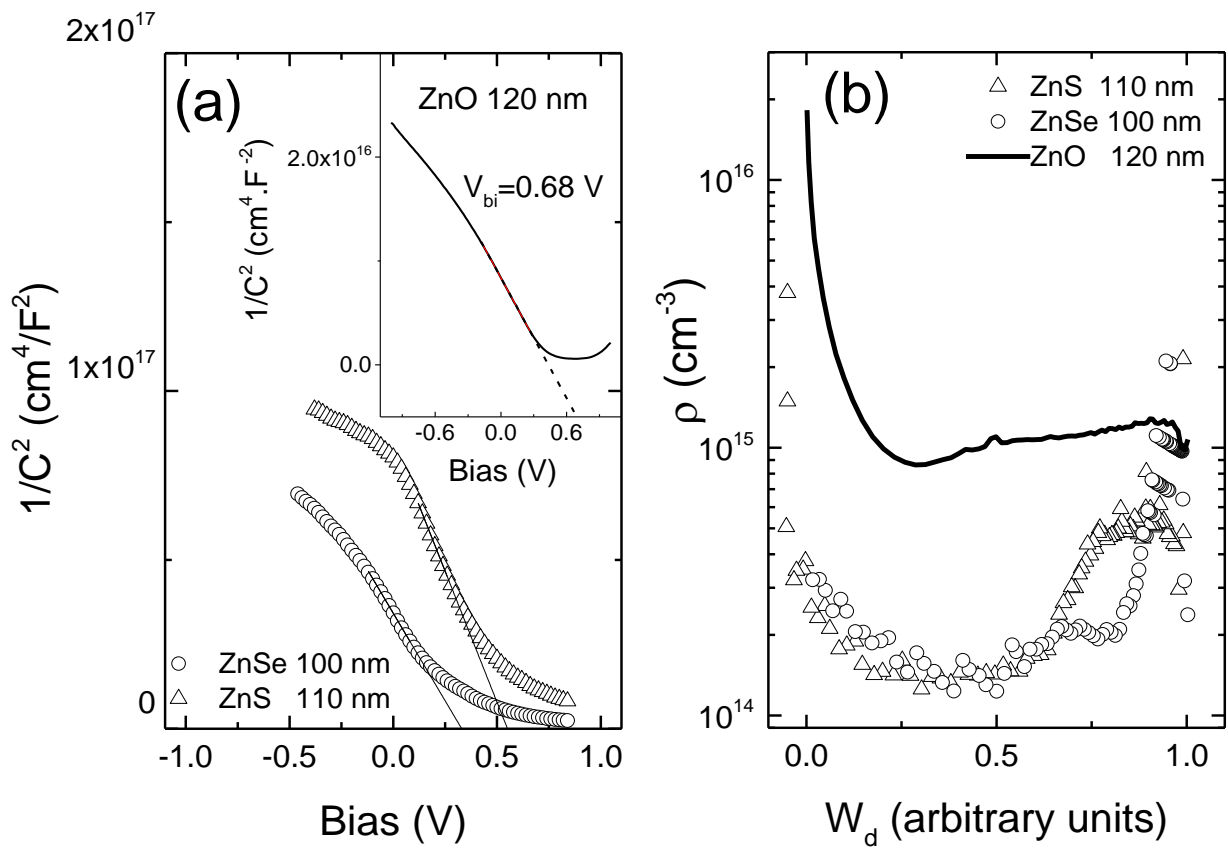

FIG. 2.

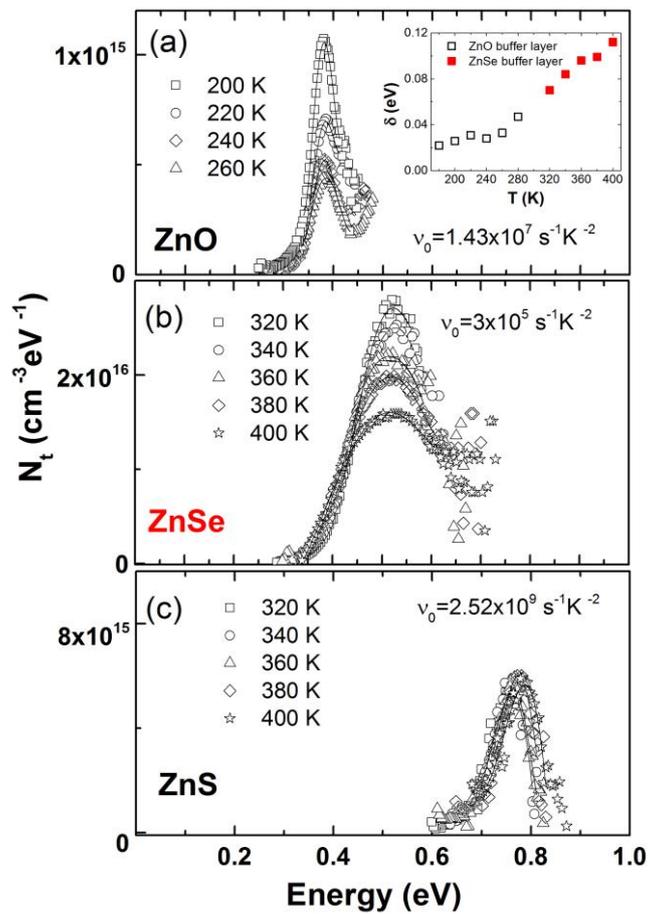

FIG. 3.

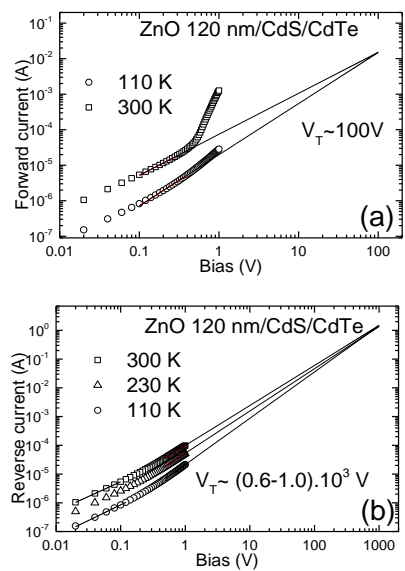

FIG. 4.

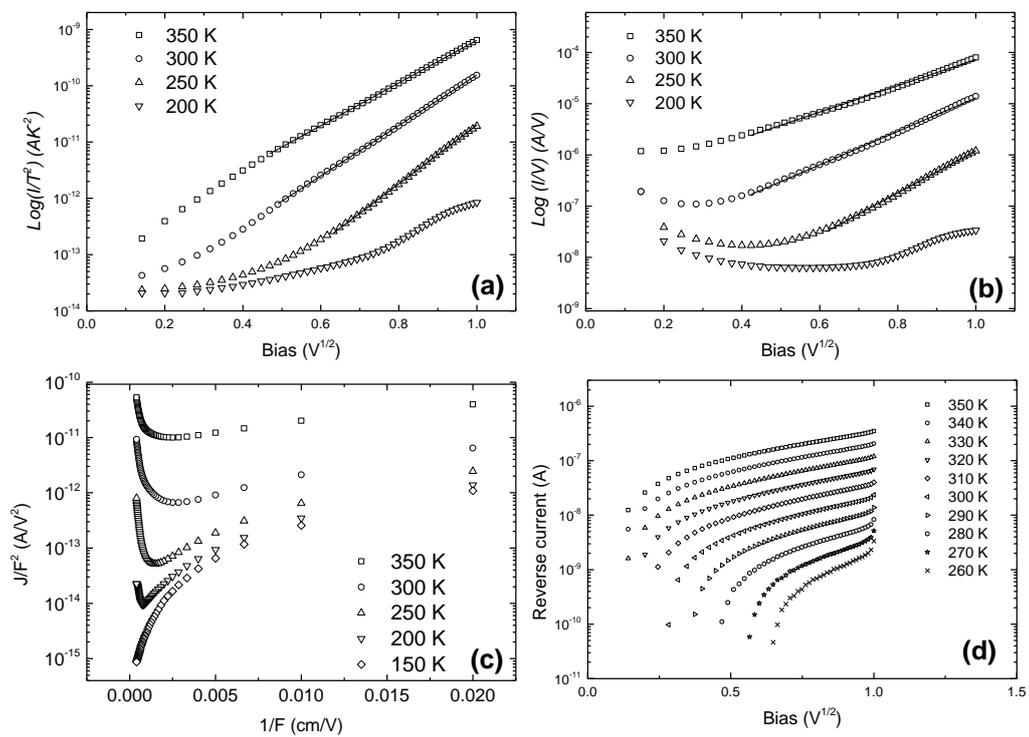

Supplemental Information (SI)

FIG. 1.

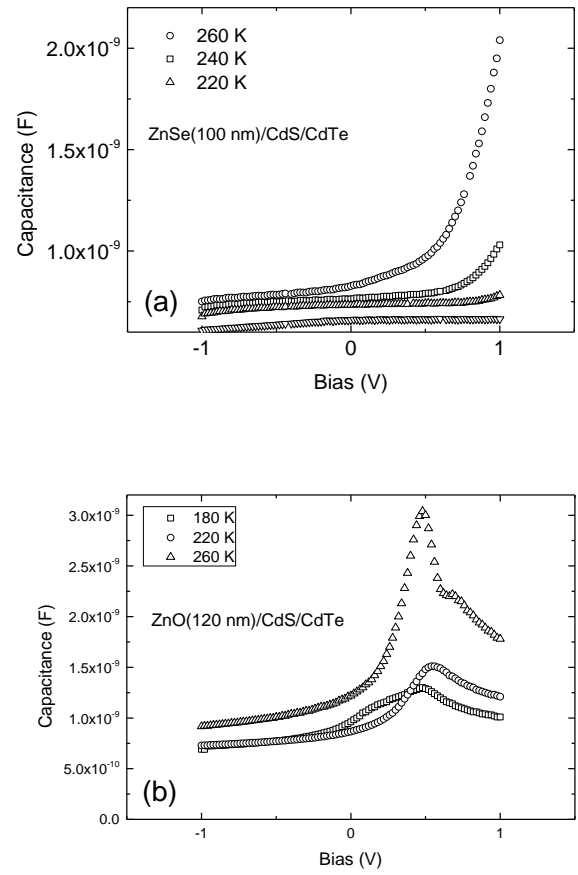

Typical *C-V* characteristics taken on the ZnSe/CdS/CdTe solar cells (a) and the ZnO/CdS/CdTe solar cells at different temperatures. The device contact area is 0.24 cm$^2$.

FIG. 2.

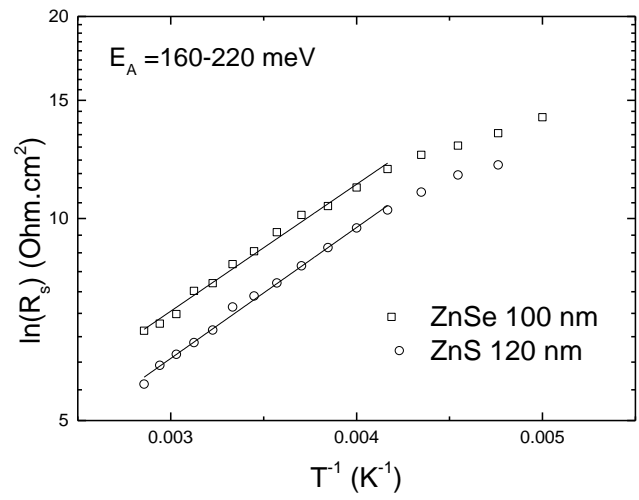

Temperature dependence of series resistance for the ZnSe/CdS/CdTe (□) and ZnS/CdS/CdTe (○) solar cells.

FIG. 3.

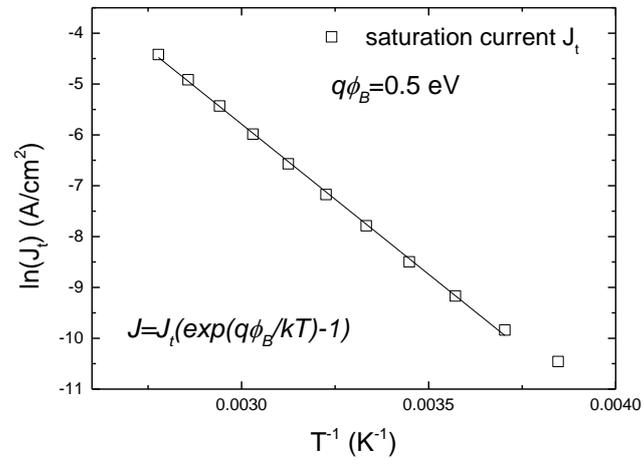

A temperature dependence of the saturation current density and the deduced barrier height at the Au contact of ZnO/CdS/CdTe solar cells.

FIG. 4.

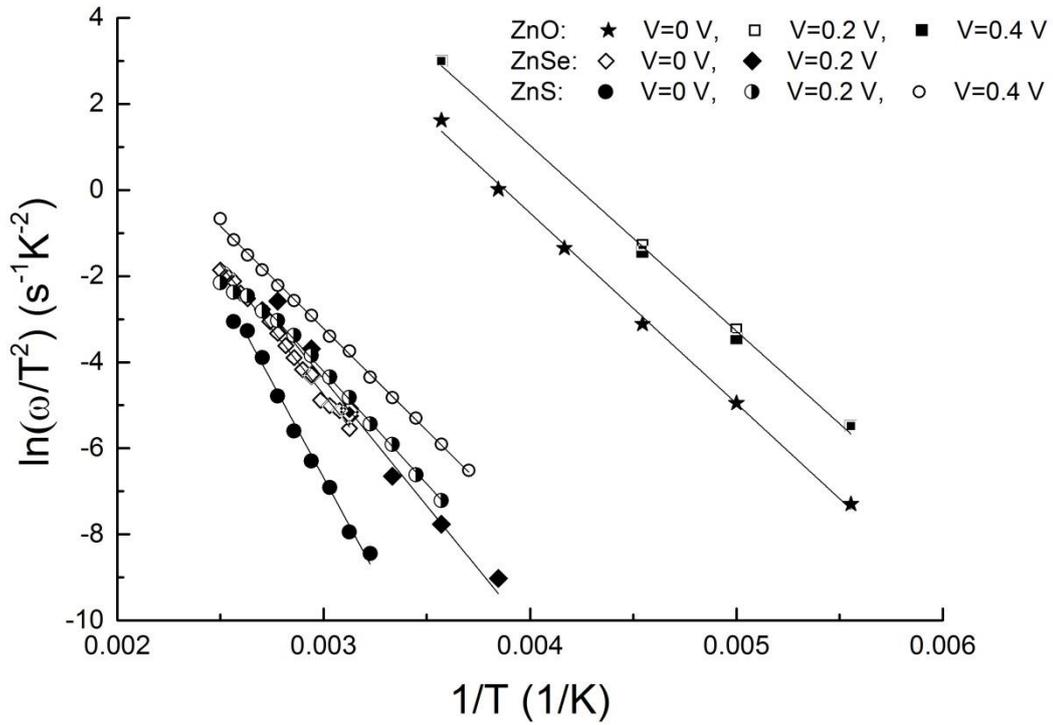

Emission rates *vs* T$^{-1}$ for the trap levels in CdTe as influenced by the choice of the buffer layer. (★(0 V), ☐ (0.2 V), ■ (0.4 V)) – ZnO, ( ◇ (0 V), ◆ (0.2 V)) - ZnSe, (●(0 V), ◐(0.2 V), ○(0.4 V)) -ZnS.

The activation energy $E_A$ did not change under different bias conditions when the CdS/CdTe interfaces were formed on ZnO indicating uniform defect distribution within the junction region. The use of ZnS buffer layers revealed bias-dependent behaviour of $E_A$. For the ZnS/CdS/CdTe structures, at V=(0.2-0.4) V the $E_A$ values decreased to 0.4-0.43 eV. In this work $E_A$ values deduced from the AS data taken at zero bias are considered.

FIG. 5.

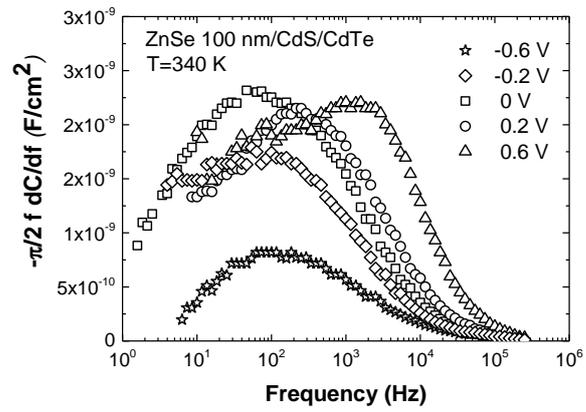

Typical capacitance derivative spectra for the ZnSe/CdS/CdTe solar cells taken at different biases applied to the back Au contact.